\documentclass[12pt]{iopart}
\usepackage{latexsym}
\usepackage{amssymb}
\usepackage{verbatim}
\usepackage{graphicx}
\usepackage{dcolumn}
\usepackage{bm}
\usepackage{setspace}
\usepackage{url}

\newcommand{\AJP}{{\it Am.~J.~Phys.~}}

\newcommand{\EPL}{{\it Europhys.~Lett.~}}

\newcommand{\MP}{{\it Mol.~Phys.~}}

\begin{document}

\title[Exploring Fluctuations and Demixing via Monte Carlo Simulation]
{Exploring Fluctuations and Phase Equilibria in \\
Fluid Mixtures via Monte Carlo Simulation} 

\author{Alan R Denton\footnote[1]{Corresponding author.
Electronic address: {\tt alan.denton@ndsu.edu}} and Michael P Schmidt}
\address{Department of Physics, North Dakota State University,
Fargo, ND 58108-6050, USA}

\date{\today}

\begin{abstract}
Monte Carlo simulation provides a powerful tool for understanding and exploring 
thermodynamic phase equilibria in many-particle interacting systems.  Among the 
most physically intuitive simulation methods is Gibbs ensemble Monte Carlo (GEMC), 
which allows direct computation of phase coexistence curves of model fluids by
assigning each phase to its own simulation cell.  When one or both of the phases 
can be modeled virtually via an analytic free energy function
[M.~Mehta and D.~A.~Kofke, \MP {\bf 79}, 39 (1993)], the GEMC method takes on
new pedagogical significance as an efficient means of analyzing fluctuations 
and illuminating the statistical foundation of phase behavior in finite systems.
Here we extend this virtual GEMC method to binary fluid mixtures and demonstrate 
its implementation and instructional value with two applications:
(1) a lattice model of simple mixtures and polymer blends and 
(2) a free-volume model of a complex mixture of colloids and polymers.  
We present algorithms for performing Monte Carlo trial moves in the virtual
Gibbs ensemble, validate the method by computing fluid demixing phase diagrams,
and analyze the dependence of fluctuations on system size.  Our open-source 
simulation programs, coded in the platform-independent Java language, are 
suitable for use in classroom, tutorial, or computational laboratory settings.
\end{abstract}

\pacs{64.60.De, 61.20.Ja, 61.20.Gy, 64.75.Xc, 01.40.-d}

\maketitle
\newpage

\section{Introduction}\label{Introduction}
Phase transitions and critical phenomena are topics fundamental to most undergraduate 
and graduate courses in thermodynamics and statistical mechanics~\cite{chandler}.
Aside from their intrinsic interest and practical relevance, phase transitions 
provide a rich conceptual context for charting the path from a microscopic Hamiltonian 
to macroscopic properties of materials via partition functions and free energies.  
Over the past sixty years, molecular simulations have provided important insights 
into the thermodynamic phase behavior of systems ranging from simple atomic fluids 
to complex macromolecular materials.  Monte Carlo and molecular dynamics methods,
in particular, have clarified the subtle interplay between energy and entropy in 
determining stability of competing phases~\cite{frenkel01,allen87}.  Monte Carlo
simulation has been further exploited as an aid in teaching statistical 
mechanics~\cite{landau73,sauer81,wilding01}.

Computer simulations of many-particle systems can readily identify mechanically 
stable (metastable) states, corresponding to local minima in the free energy of 
the system as a function of externally controlled parameters.  For fluid systems, 
external parameters may include temperature, pressure, density, and (in the case 
of mixtures) concentration.  More difficult is finding the {\it global} minimum
in the free energy, which is required to establish {\it thermodynamic} stability.
The task is especially challenging near a first-order phase transition, where 
two bulk phases (e.g., vapor and liquid in a one-component system, or $A$-rich
and $B$-rich phases in a mixture of $A$ and $B$ species) may coexist in equilibrium.

In simulations of finite model systems, the free energy cost associated with 
interfaces between phases results in hysteresis and tends to inhibit the simultaneous
presence of more than one phase in a single simulation cell~\cite{frenkel01}.
For this reason, mapping out thermodynamic phase diagrams using computer simulation
traditionally involves computing the free energy, usually via thermodynamic 
integration of the internal energy, and then performing a coexistence analysis 
via a Maxwell common-tangent construction.  An alternative approach, which has been
applied to both fluid and magnetic systems, is histogram reweighting~\cite{wilding01}.

A more direct and intuitive route to analyzing phase equilibria and calculating densities 
of coexisting phases is the Gibbs ensemble Monte Carlo (GEMC) method.  Introduced by 
Panagiotopoulos~\cite{panagiotopoulos87,panagiotopoulos88,panagiotopoulos92,
panagiotopoulos95}, this method models each phase in its own simulation cell.
By elegantly avoiding both the complication of interfaces and the need for 
thermodynamic integration, the GEMC method provides a computationally efficient 
and conceptually transparent means of computing fluid phase diagrams.  
The GEMC method has been widely applied as a research tool to analyze phase behavior
of simple fluids and fluid mixtures~\cite{panagiotopoulos88}, as well as complex fluids,
such as colloid-polymer mixtures~\cite{panagiotopoulos95,bolhuis02,lu-denton11} 
and charge-stabilized colloidal suspensions~\cite{lu-denton07}.  
To date, however, the pedagogical potential of the GEMC method has been 
relatively less appreciated.  

In this paper, we explore the GEMC method as a tool for introducing students to the
statistical nature of fluctuations and phase behavior in finite systems.  We are 
inspired by a useful variation of the method, proposed by Mehta and Kofke~\cite{kofke93},
which incorporates a thermodynamic model via an analytic free energy or equation 
of state (EOS).  By modeling one of the phases explicitly and the other as
virtual (via an EOS), the GEMC-EOS method gains efficiency over the original 
GEMC method and is reliable when the EOS of the virtual phase is accurately known.

The GEMC-EOS method was conceived and tested for one-component systems and applied 
to a simple fluid of particles interacting via Lennard-Jones pair potentials.  
A simplified, pedagogically appealing version of the GEMC-EOS method models 
{\it both} phases as virtual~\cite{kofke93}.  The purpose of the present work is 
to extend this ``virtual GEMC method" to fluid mixtures and to demonstrate the value
of the method in exploring and elucidating the statistical nature of demixing.

The remainder of the paper is organized as follows.  In Sec.~\ref{Methods}, we 
briefly review the GEMC-EOS method and then describe our extension to fluid mixtures.
In Sec.~\ref{Applications}, we demonstrate the application of the virtual GEMC method 
to the demixing behavior of two model systems: (1) a lattice model of mixtures,
and (2) a free-volume model of colloid-polymer mixtures.  For the latter,
we analyze density fluctuations as functions of system size and proximity to
the critical point.
Finally, in Sec.~\ref{Conclusions}, we conclude by emphasizing the significance 
of the virtual GEMC method as a classroom and computational laboratory tool.

\section{Methods}\label{Methods}
\subsection{Gibbs Ensemble Monte Carlo Simulation}\label{gemc}
\begin{figure}
\centering\includegraphics[width=0.6\textwidth,angle=0]{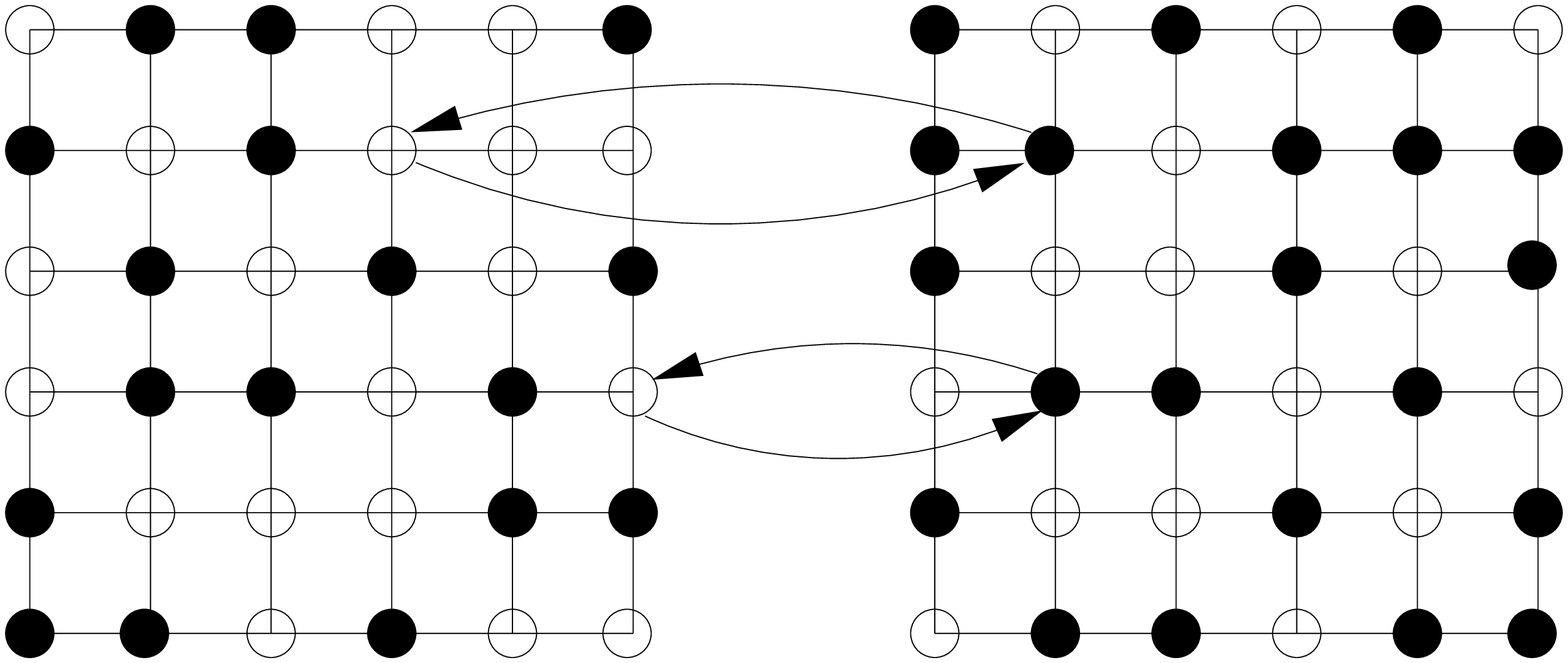}
\\[4ex]
\centering\includegraphics[width=0.6\textwidth,angle=0]{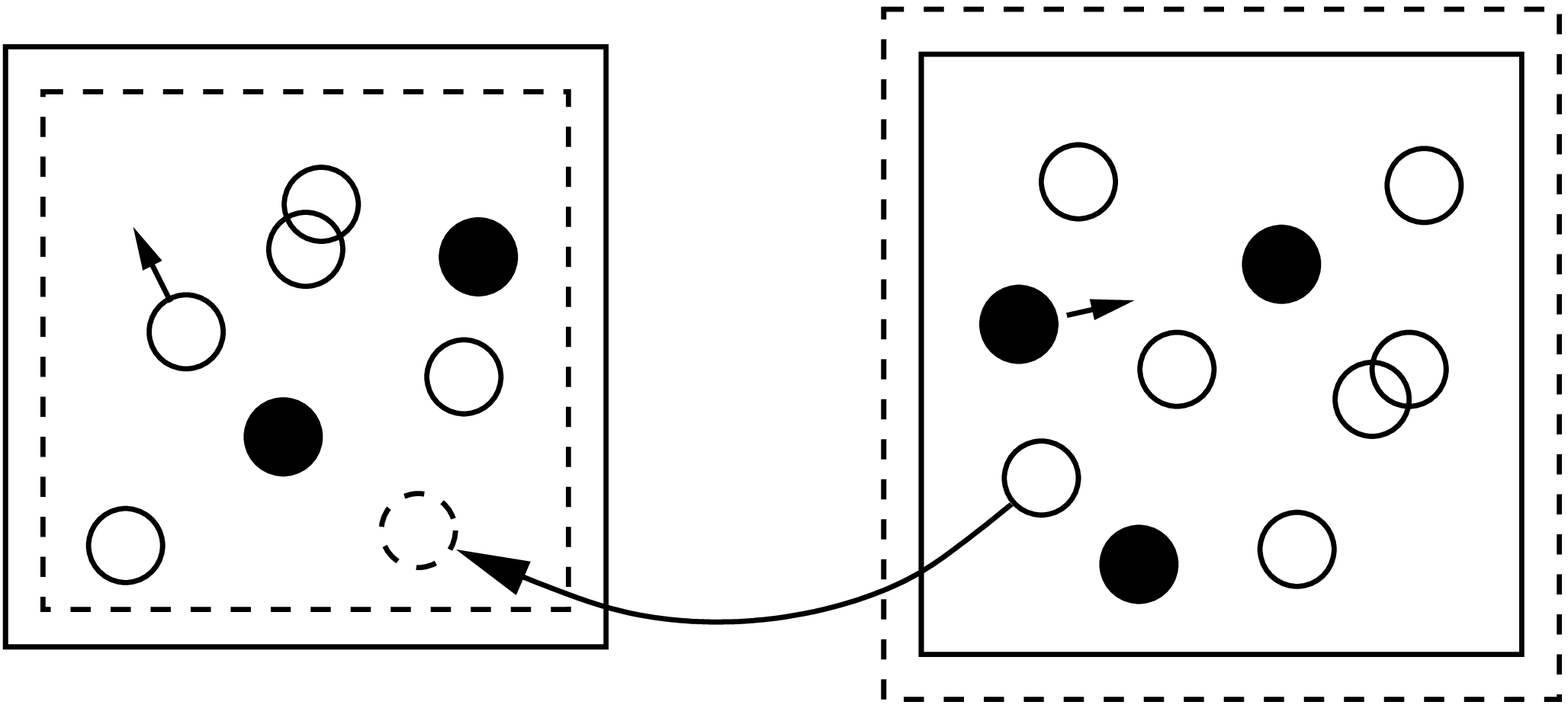}
\caption{Schematic illustration of trial moves performed in the Gibbs ensemble 
Monte Carlo method.  Top: particle transfers between two boxes in the lattice model 
of binary mixtures (Sec.~\ref{lattice-model}).  Bottom:  particle displacements, 
volume changes, and particle transfers in a model of colloid-polymer mixtures 
(Sec.~\ref{cp-mixtures}).  The virtual GEMC method replaces both of the boxes 
by a virtual phase described by a free energy.
}
\label{figure1}
\end{figure}

By allowing distinct simulation boxes to exchange particles and volume, but not 
otherwise interact, the Gibbs ensemble Monte Carlo method can efficiently
equilibrate coexisting phases (see Fig.~\ref{figure1}).  Each box accommodates 
one of the phases at equal temperatures, pressures, and chemical potentials,
with no need for interfaces.
Originally introduced by Panagiotopoulos~\cite{panagiotopoulos87} to model liquid-vapor 
coexistence of simple one-component fluids, the GEMC method was later extended to 
mixtures~\cite{panagiotopoulos88}.
Since its introduction, the GEMC method has been refined and used to map out phase diagrams 
for a wide variety of systems~\cite{panagiotopoulos92,panagiotopoulos95}.

Trial moves in Monte Carlo simulations of thermal systems are typically accepted with 
probabilities consistent with the condition of detailed balance~\cite{frenkel01}.
According to this condition, the average rate of transition from an old ($o$) to a 
new ($n$) state equals, in equilibrium, the average reverse transition rate:
\begin{equation}
P(o)\pi(o \to n)=P(n)\pi(n \to o)~,
\label{db}
\end{equation}
where $P(o)$ and $P(n)$ are probabilities of finding the system in states $o$ and $n$
and $\pi(o \to n)$ is the transition probability between $o$ and $n$.
Assuming that transitions $o \to n$ and $n \to o$ are attempted at equal rates,
the classic Metropolis algorithm~\cite{metropolis53} imposes Eq.~(\ref{db}) by
accepting trial moves with probability~\cite{frenkel01,allen87}
\begin{equation}
{\cal P}(o \to n) = \min\left\{1,~\frac{P(n)}{P(o)}\right\}~.
\label{Metropolis}
\end{equation}

In the canonical (constant-$NVT$) Gibbs ensemble, a microstate of a binary mixture,
at absolute temperature $T$, is specified by the positions of all particles 
(collectively denoted by $\{{\bf r}\}$), the volume $V_i$ of each box ($i=1,2$), 
under the constraint of constant total volume $V=V_1+V_2$, and the particle numbers 
$N_{ij}$ of the two species ($j=A, B$) in each box, with constant total numbers
$N_1=N_{A1}+N_{B1}$ and $N_2=N_{A2}+N_{B2}$.  
The probability distribution for the possible microstates is given by
\begin{equation}
P(\{N_{ij}\}, \{V_i\}, T; \{{\bf r}\})\propto P_{\rm id}
~e^{-\beta U(\{{\bf r}\})}~,
\label{P}
\end{equation}
where $\beta\equiv 1/(k_BT)$, $U=U_1+U_2$ is the internal energy of
both boxes, $U_i$ being the internal energy of box $i$, and
\begin{equation}
P_{\rm id}(\{N_{ij}\}, \{V_i\})\propto 
\frac{V_1^{N_1}V_2^{N_2}}{N_{A1}!N_{B1}!N_{A2}!N_{B2}!}
\label{Pid}
\end{equation}
is the probability distribution for an ideal mixture.

\subsection{Virtual Gibbs Ensemble Monte Carlo: Modeling Phases by a Free Energy}\label{eos-gemc}
The GEMC-EOS method replaces one of the simulation boxes by a {\it virtual} phase 
containing no explicit particles, but represented instead by a thermodynamic model 
in the form of a free energy or equation of state. 
In the virtual GEMC method, {\it both} phases are modeled as virtual with a prescribed
excess free energy $F_{\rm ex}$, associated with interparticle interactions.
The probability density for a configuration in which box $i$ has volume $V_i$
and contains $N_{ij}$ particles of species $j$ is then given by 
[{\it cf}.~Eq.~(\ref{P})]
\begin{equation}
P\propto P_{\rm id}~e^{-\beta F_{\rm ex}}~.
\label{P-eos}
\end{equation}

According to Eqs.~(\ref{Metropolis}) and (\ref{P-eos}), the acceptance probability 
for a trial move from an old ($o$) to a new ($n$) state is given by the ratio 
of the corresponding probability densities:
\begin{equation}
{\cal P}(o \to n) = \min\left\{1, \frac{P_{\rm id}(n)}{P_{\rm id}(o)}
~e^{-\beta\Delta F_{\rm ex}}\right\}~,
\label{acc}
\end{equation}
where $\Delta F_{\rm ex}=F_{\rm ex}(n)-F_{\rm ex}(o)$ is the associated change in 
excess free energy.
From Eq.~(\ref{Pid}), a trial transfer of volume $\Delta V$ from phase 1 to phase 2 
($V_1\to V_1-\Delta V$, $V_2\to V_2+\Delta V$) is accepted with minimum probability
\begin{equation}
{\cal P}_{\rm vol}=
\left(1-\frac{\Delta V}{V_1}\right)^{N_1}\left(1+\frac{\Delta V}{V_2}\right)^{N_2}
~e^{-\beta\Delta F_{\rm ex}}~.
\label{acc-volume}
\end{equation}
In practice, trial moves in $\ln(V_1/V_2)$ improve efficiency, 
with acceptance probability~\cite{frenkel01,allen87}
\begin{equation}
{\cal P}'_{\rm vol} =
\left(1+\frac{\Delta V}{V_1}\right)^{N_1+1}
\left(1-\frac{\Delta V}{V_2}\right)^{N_2+1} e^{-\beta\Delta F_{\rm ex}}~.
\label{acc-vol2}
\end{equation}
Finally, a trial transfer of a particle, say of species $A$ from phase 1 to 2 
($N_{A1}\to N_{A1}-1$, $N_{A2}\to N_{A2}+1$) is accepted with minimum probability
\begin{equation}
{\cal P}_{\rm trans} = 
\frac{V_2}{V_1}~\frac{N_{A1}}{N_{A2}+1}
~e^{-\beta\Delta F_{\rm ex}}~.
\label{acc-transfer}
\end{equation}
In the next section, to illustrate the utility of the virtual GEMC method in 
exploring demixing phenomena, we apply the method to two model mixtures.

\section{Applications of the Virtual GEMC Method to Demixing}\label{Applications}
\subsection{Lattice Model of Binary Mixtures}\label{lattice-model}
A simple lattice model provides an instructive introduction to phase separation. 
Consider a lattice of $N$ sites, each occupied by a particle of type $A$ or $B$ 
with volume fractions defined as $\phi_A\equiv N_A/N$ and $\phi_B\equiv N_B/N$, the 
total volume fraction being conserved: $\phi_A+\phi_B=1$ (see Fig.~\ref{figure1}).
The entropy of mixing is given by
\begin{equation}
S_{\rm mix}=k_B\ln\left(\frac{N!}{N_A!N_B!}\right)
\simeq -k_b\left[N_A\ln\phi_A+N_B\ln\phi_B\right]~,
\label{Smix}
\end{equation}
using Stirling's approximation, $\ln N!\simeq N\ln N -N$ (valid for $N\gg 1$).
Suppose now that only nearest-neighbor pairs interact and that correlations 
between particles can be neglected.  In this mean-field approximation, 
a particle of type $i$ occupies a given site with probability $\phi_i$.  
The internal energy, then independent of configuration, can be expressed as
\begin{equation}
U=\frac{1}{2}\left(N_A\phi_A\epsilon_{AA}+N_B\phi_B\epsilon_{BB}\right)
+N_A\phi_B\epsilon_{AB}~,
\label{U}
\end{equation}
where $\epsilon_{ij}$ denotes the interaction energy between a pair of particles 
of species $i$ and $j$.  The corresponding internal energy of mixing is 
\begin{equation}
U_{\rm mix}=U-\frac{1}{2}\left(N_A\epsilon_{AA}+N_B\epsilon_{BB}\right)
=\chi N_A\phi_B~,
\label{Umix}
\end{equation}
where $\chi\equiv\epsilon_{AB}-(\epsilon_{AA}+\epsilon_{BB})/2$ is the Flory 
interaction parameter, which characterizes the incompatibility of the two species.
Finally, with $\phi\equiv\phi_A$, the Helmholtz mixing free energy, 
$F_{\rm mix}\equiv U_{\rm mix}-TS_{\rm mix}$, is approximated (per site) by
\begin{equation}
\frac{\beta F_{\rm mix}}{N}
=\phi\ln\phi+(1-\phi)\ln(1-\phi)+\chi\phi(1-\phi)~.
\label{Fmix}
\end{equation}
\begin{figure}
\centering\includegraphics[width=0.6\textwidth,angle=0]{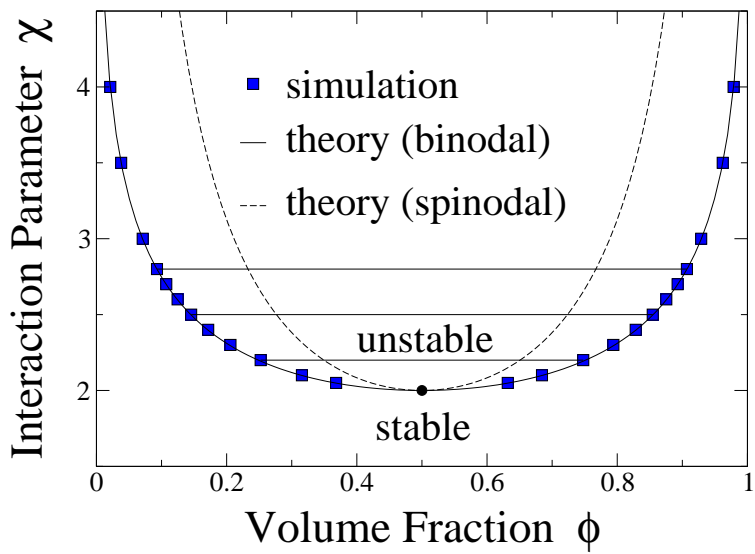}
\caption{Phase diagram for the lattice mixture model.  
Squares: data from virtual GEMC simulations with both phases described by 
the mixing free energy of Eq.~(\ref{Fmix}).
Solid and dashed curves: exact binodal and spinodal curves 
[Eqs.~(\ref{binodal}), (\ref{spinodal})].
Circle: theoretical critical point.
Representative tie lines join coexisting phases on the binodal.
}
\label{figure2}
\end{figure}

The phase behavior of the lattice model predicted by the above mean-field theory
is easily deduced from the analytic expression for the free energy [Eq.~(\ref{Fmix}].
In equilibrium, two phases coexist at equal temperatures, pressures, and 
chemical potentials, $\mu_A$ and $\mu_B$, of the two species.  
Incompressibility of the lattice and conservation of total volume fraction 
reduce the coexistence criteria to a single condition: 
$\mu_A=\partial F_{\rm mix}/\partial\phi=0$.
The coexistence curve (binodal) thus takes the analytic form
\begin{equation}
\chi=\frac{1}{2\phi-1}\ln\left(\frac{\phi}{1-\phi}\right)~.
\label{binodal}
\end{equation}
The inflection points of the free energy define the spinodal curve,
\begin{equation}
\frac{\partial^2 F_{\rm mix}}{\partial\phi^2}=0
\quad\Rightarrow\quad 
\chi~=~\frac{1}{2\phi(1-\phi)}~,
\label{spinodal}
\end{equation}
inside of which the mixture is thermodynamically unstable and spontaneously demixes.
The binodal and spinodal curves terminate at a lower critical point, 
$\phi_{\rm crit}=0.5$, $\chi_{\rm crit}=2$, below which the mixture is stable.
Figure~\ref{figure2} shows the resulting phase diagram, where the 
interaction parameter $\chi$ plays the role of an inverse temperature.

In the virtual GEMC method, each phase is governed by the mean-field free energy 
of Eq.~(\ref{Fmix}).  With total volume fraction conserved, the only independent 
trial moves are exchanges of particles between the two boxes.
A trial exchange of an $A$ particle in box 1 for a $B$ particle in box 2,
for example, is accepted with probability
\begin{equation}
{\cal P}_{\rm exch}=e^{-\beta\Delta F_{\rm mix}}
=\frac{N_{A1}N_{B2}}{(N_{A2}+1)(N_{B1}+1)}~e^{-\beta\Delta U_{\rm mix}}~,
\label{acc-mix-transfer1}
\end{equation}
where $F_{\rm mix}$ is the total mixing free energy and
\begin{equation}
\Delta U_{\rm mix}=2\chi(\phi_1-\phi_2-1/N_1)
\label{delta-Umix}
\end{equation}
with $\phi_i\equiv N_{Ai}/N_i$.
For sufficiently large systems, Eq.~(\ref{acc-mix-transfer1}) can be approximated by 
\begin{equation}
{\cal P}_{\rm exch}
=\frac{\phi_1(1-\phi_2)}{\phi_2(1-\phi_1)}
~e^{-\beta\Delta U_{\rm mix}}~.
\label{acc-mix-transfer2}
\end{equation}
The Boltzmann factor in Eqs.~(\ref{acc-mix-transfer1}) and (\ref{acc-mix-transfer2}) 
evidently favors demixing if $\chi>0$, while the entropic prefactor always favors 
mixing.  The competition is decided by the magnitude of $\chi$.

As an illustration of the GEMC method with both phases treated virtually, 
we have implemented Eq.~(\ref{acc-mix-transfer2}) for the lattice model and 
performed simulations to calculate several points on the coexistence curve (binodal).  
A trial move figuratively flips a coin (i.e., generates a random number)
to decide, with equal probabilities, whether to exchange an $A$ particle for a $B$
particle, or a $B$ for an $A$, between the two boxes.  Choosing $N_1=N_2=1000$, 
and initializing with equal numbers of $A$ and $B$ particles in each box,
we equilibrated the system for $10^4$ MC steps, a step being defined as a trial exchange
of every particle.  We then accumulated data over the next $10^4$ steps to calculate 
average values of $\phi_{A1}$ and $\phi_{A2}$ over a range of $\chi$ values.  
Because virtual phases have no explicit particles to displace, these simulations are 
extremely fast, with typical CPU times of a few minutes on a PC, scaling linearly 
with system size.  

Our numerical results, plotted in Fig.~\ref{figure2}, agree essentially exactly
with the analytic expression for the binodal [Eq.~(\ref{binodal})], thus validating
the virtual GEMC method.  On approaching the critical point, we observed growth 
of fluctuations and frequent switching of phases between the two boxes.  
Phase switching is easily suppressed, however, by increasing the number of particles.  
In a tutorial setting, students could, for example, probe miscibility 
in different regions of the phase diagram and explore the dependence of fluctuations 
on system size and proximity to the critical point.  For this purpose, it is valuable
that the binodal and spinodal curves are also known analytically.

\begin{figure}
\centering\includegraphics[width=0.6\textwidth,angle=0]{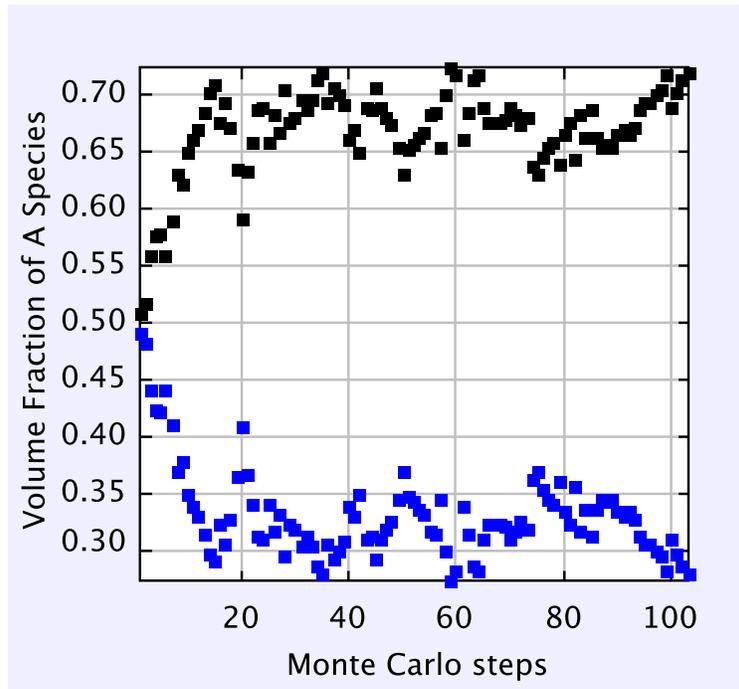}
\caption{Typical diagnostic data from a virtual GEMC simulation of the 
lattice mixture model [Eq.~(\ref{Fmix})]: cumulative average volume fraction of 
$A$ species in each box vs.~number of Monte Carlo steps during equilibration for 
interaction parameter $\chi=2.1$.
}
\label{figure3}
\end{figure}

To facilitate pedagogical applications on a variety of platforms, we have coded our 
simulations in the Java programming language using the Open Source Physics (OSP) 
library~\cite{osp,download}.  The graphical user interface provided by the 
{\tt AbstractSimulation} class in the {\tt controls} package of the OSP library
allows the user to easily input parameters and start, stop, and step through a 
simulation.  Figure~\ref{figure3} shows typical diagnostic data, collected during 
the equilibration stage, displayed with the OSP {\tt frames} package.  
Our code can be easily made compatible, however, with any graphics library.

A closely related model depicts a polymer blend as a mixture of chains of lengths 
(degrees of polymerization) $M_A$ and $M_B$, whose segments (monomers) fully occupy 
the sites of a lattice.  Connecting monomers to form chains reduces their entropic
free energy by a factor of inverse chain length.  Making an analogous mean-field 
approximation, the Flory-Huggins theory for this model yields~\cite{degennes79}
\begin{equation}
\frac{\beta F_{\rm mix}}{N}=\frac{\phi}{M_A}\ln\phi
+\frac{1-\phi}{M_B}\ln(1-\phi)+\chi\phi(1-\phi)~,
\label{FH}
\end{equation}
where $N_A$ now represents the number of $A$ chains (rather than $A$ segments)
and $\phi=N_AM_A/N$.
In the symmetric case ($M_A=M_B=M$), the phase diagram is identical to Fig.~\ref{figure2},
but with $\chi$ replaced by $M\chi$ on the vertical axis.
Asymmetric blends ($M_A\neq M_B$) display much richer phase behavior.
The Flory-Huggins mixing free energy [Eq.~(\ref{FH})] could be used to
explore, via the virtual GEMC method, demixing of polymer blends.

\subsection{Colloid-Polymer Mixtures}\label{cp-mixtures}
As a second application of the virtual GEMC method, we consider a widely studied
system from soft matter physics -- a suspension of colloidal particles mixed with 
free (nonadsorbing) polymers~\cite{pusey,poon02}.
The classic Asakura-Oosawa-Vrij (AOV) model~\cite{ao54,vrij76} idealizes the 
colloidal particles as hard spheres, monodisperse in radius $R_c$, and the 
polymer coils as effective spheres with a radius $R_p$ equal to the average
radius of gyration.  The polymers are modeled as mutually ideal (noninteracting), 
but impenetrable to the colloids, with which they have hard-sphere interactions.
Although real polymer coils fluctuate in size~\cite{denton-schmidt02}, the AOV model 
portrays the effective polymer spheres as monodisperse in size (see Fig.~\ref{figure1}).
The size ratio $q=R_p/R_c$ is then the one model parameter that distinguishes 
different mixtures.

The thermodynamic state of the system is specified by the total volume $V$ and 
numbers of colloids and polymers, $N_c$ and $N_p$, with respective number densities
$\rho_c=N_c/V$ and $\rho_p=N_p/V$ and volume fractions $\phi_c=(4\pi/3)\rho_c R_c^3$
and $\phi_p=(4\pi/3)\rho_p R_p^3$.  In the Gibbs ensemble, the particle numbers 
in the two boxes are denoted $N_{c1}$, $N_{c2}$, $N_{p1}$, $N_{p2}$, and constrained
by $N_{c1}+N_{c2}=N_c$ and $N_{p1}+N_{p2}=N_p$.  With only hard interparticle interactions, 
the thermodynamic state is independent of temperature, there being no energy scale.
For contact with experiments, it is helpful to imagine the system in osmotic equilibrium
with a reservoir of pure polymer, which exchanges polymers with the system 
to maintain fixed polymer chemical potential.
The reservoir density $\rho_p^r$ plays the role of an inverse temperature.

To describe the phase behavior of the AOV model of colloid-polymer mixtures, 
Lekkerkerker \etal~\cite{lekkerkerker92} have developed a free-volume theory 
by expressing the Helmholtz free energy density (to within a constant) as
\begin{equation}
f(\phi_c,\phi_p) = k_BT\rho_c\left(\ln\phi_c-1\right)+f_{hs}(\phi_c)
+f_p(\phi_c,\phi_p)~.
\label{ffv1}
\end{equation}
The first two terms on the right side are the colloid ideal-gas and hard-sphere 
excess free energy densities, the latter being accurately approximated by 
the Carnahan-Starling relation~\cite{hansen90}:
\begin{equation}
\beta f_{hs}(\phi_c)=\rho_c\frac{\phi_c(4-3\phi_c)}{(1-\phi_c)^2}~.
\label{fhs}
\end{equation}
If colloid-polymer correlations are neglected (mean-field approximation), 
the polymer free energy density can be approximated by that of an ideal gas 
of polymers confined to the {\it free} volume, i.e., 
the volume not excluded by the hard-sphere colloids:
\begin{equation}
f_p(\phi_c,\phi_p) = k_BT\rho_p\left[\ln\left(\frac{\phi_p}{\alpha(\phi_c)}\right)-1\right]~.
\label{fp1}
\end{equation}
Here $\alpha(\phi_c)$ is the free-volume fraction of polymers amidst colloids,
which is reasonably approximated by scaled-particle theory:
\begin{equation}
\alpha(\phi_c) =
\frac{1}{1-\phi_c}\exp\left(-\sum_{m=1}^3 C_m \gamma^m\right)~,
\label{alpha}
\end{equation}
where $\gamma=\phi_c/(1-\phi_c), C_1=3q+3q^2+q^3, C_2=(9q^2/2)+3q^3$, 
and $C_3=3q^3$.  For ideal polymers, equality of polymer chemical potentials 
in the system and reservoir implies $\phi_p=\phi_p^r\alpha(\phi_c)$, where
$\phi_p^r=(4\pi/3)\rho_p^r R_p^3$ is the volume fraction of the polymer reservoir.

\begin{figure}
\centering\includegraphics[width=0.6\textwidth,angle=0]{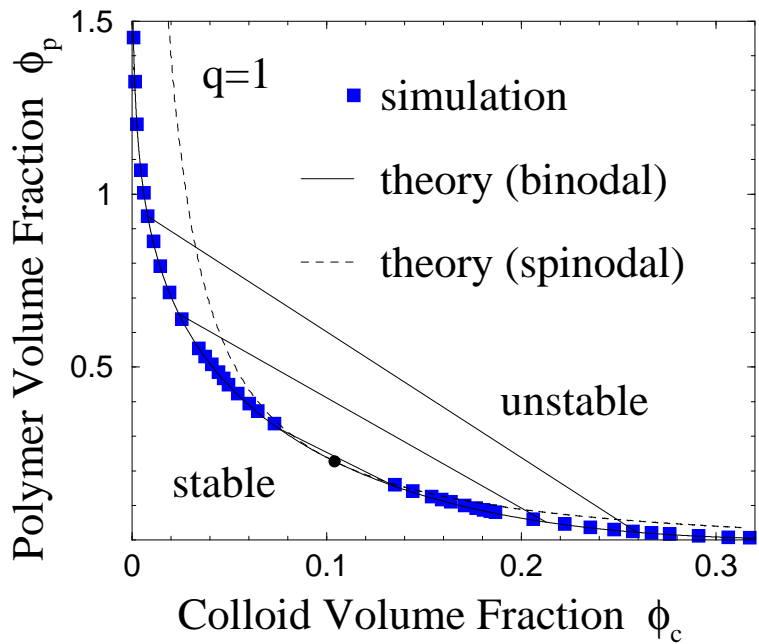}
\caption{Phase diagram for the AOV model of colloid-polymer mixtures 
with size ratio (polymer radius of gyration over colloid radius) $q=1$.  
Squares: data from virtual GEMC simulations with both phases described by 
free energy from free-volume theory.  Solid and dashed curves: binodal and 
spinodal curves from Maxwell common-tangent construction.
Circle: theoretical critical point.
Tie lines join coexisting phases on the binodal.
}
\label{figure4}
\end{figure}
Given the analytic expression for the free energy [Eqs.~(\ref{ffv1})-(\ref{alpha})],
it is straightforward to explore phase stability of the AOV model predicted by 
the mean-field free-volume theory.  The demixing binodal can be calculated from
a Maxwell construction that equates pressures and chemical potentials of coexisting 
phases, while the spinodal is defined by the inflection points of the free energy.  
Figure~\ref{figure4} shows the resulting fluid demixing phase diagram for mixtures
with a size ratio $q=1$, for which the polymer coils are impenetrable to the colloids.
The colloid-rich and colloid-poor branches of the binodal correspond, respectively, 
to colloidal ``liquid" and ``vapor" phases.

For simulations, the free energy expressed by Eqs.~(\ref{ffv1})-(\ref{alpha})
can be recast in the form
\begin{equation}
f(\phi_c,\phi_p) = f_{\rm id}(\phi_c,\phi_p) + f_{\rm ex}(\phi_c,\phi_p)~, 
\label{ffv2}
\end{equation}
comprising an ideal-gas free energy density,
\begin{equation}
\beta f_{\rm id}(\phi_c,\phi_p) =
\rho_c\left(\ln\phi_c-1\right)+\rho_p\left(\ln\phi_p-1\right)~,
\label{fid}
\end{equation}
and an excess free energy density
\begin{equation}
\beta f_{\rm ex}(\phi_c,\phi_p) = \beta f_{hs}(\phi_c) - \rho_p\ln\alpha(\phi_c)~.
\label{fex}
\end{equation}
The acceptance probabilities for the trial moves in the virtual Gibbs ensemble are 
given by Eqs.~(\ref{acc-vol2}) and (\ref{acc-transfer}) combined with Eq.~(\ref{fex}).
A transfer of volume $\Delta V$ from phase 1 to 2 is accepted with probability
\begin{equation}
{\cal P}'_{\rm vol} =
\left(1-\frac{\Delta V}{V_1}\right)^{N_1+1}
\left(1+\frac{\Delta V}{V_2}\right)^{N_2+1} 
e^{-\beta\Delta F_{\rm ex}}~,
\label{acc-cp-vol2}
\end{equation}
while transferring a colloid from phase 1 to 2 is accepted with probability, 
\begin{equation}
{\cal P}_{\rm trans} = 
\frac{V_2}{V_1}~\frac{N_{c1}}{N_{c2}+1}
~e^{-\beta\Delta F_{\rm ex}}~.
\label{acc-cp-transfer}
\end{equation}
The last factor on the right side of Eqs.~(\ref{acc-cp-vol2}) and 
(\ref{acc-cp-transfer}) may be expressed as
\begin{equation}
e^{-\beta\Delta F_{\rm ex}} = 
\left(\frac{\alpha_1(n)}{\alpha_1(o)}\right)^{N_{p1}}
\left(\frac{\alpha_2(n)}{\alpha_2(o)}\right)^{N_{p2}}
e^{-\beta\Delta F_{\rm hs}}~.
\label{eFex}
\end{equation}

To illustrate our implementation of the virtual GEMC method for the AOV model, we have
performed simulations of mixtures with a size ratio of $q=1$.  For efficiency, we 
pre-computed $f_{hs}$ and $\alpha$ vs.~$\phi_c$ and stored these data in a lookup table.  
Initializing the system with equal numbers of particles and equal volumes 
in the two boxes, we fixed the total number of colloids at $N_c=2000$, chose
the volume to give an average colloid volume fraction of $\phi_c=0.1$, and
adjusted the polymer volume fraction by varying the total number of polymers 
over a range $2500<N_p<10^4$.
After equilibrating for $10^4$ steps, we accumulated statistics over the next 
$10^4$ steps to calculate average volume fractions of each species in each box.
A step is here defined as one trial volume exchange and a trial transfer of every 
particle.  In attempting a particle transfer, we first randomly chose a box (1 or 2),
then randomly chose a species of particle to be transferred to the other box.  

The results of our simulations are plotted on the demixing phase diagram in
Fig.~\ref{figure4}.  Each run generated a pair of points on the binodal 
-- one on the liquid branch and one on the vapor branch.  Theory and simulation
again agree very closely, further validating the method.  
As with the lattice model, students could, in a computational laboratory setting, 
explore phase stability in different parts of the phase diagram and explore the 
variation of fluctuations with system size and proximity to the critical point. 
Although the free-volume theory of the AOV model does not yield the binodal and 
spinodal curves in analytic form, the calculation of these curves by a 
Maxwell construction on the free energy is straightforward.

\begin{figure}
\centering\includegraphics[width=0.6\textwidth,angle=0]{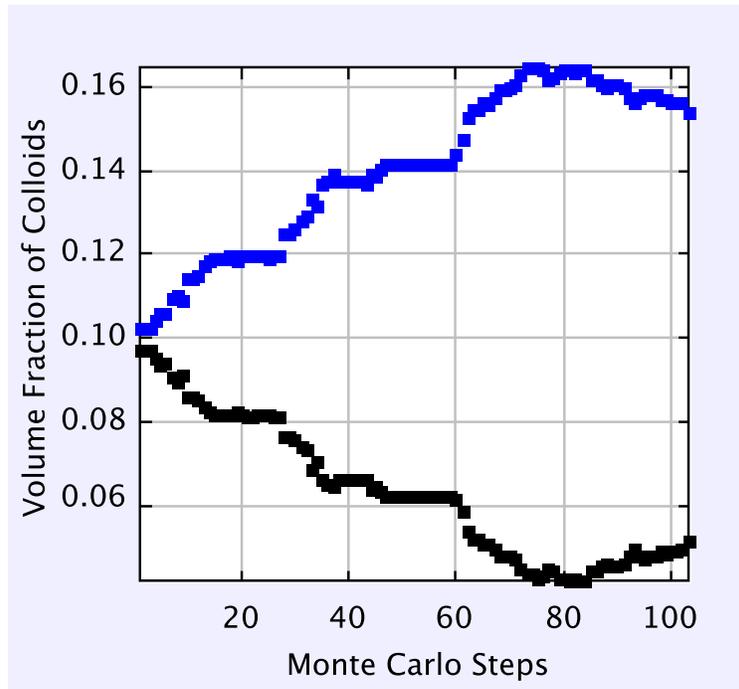}
\caption{Typical diagnostic data from virtual GEMC simulation of AOV model
of colloid-polymer mixtures: cumulative average volume fractions of colloids
in each box vs.~Monte Carlo steps during equilibration for size ratio $q=1$
and total volume fractions $\phi_c=0.1$, and $\phi_p=0.3$.
}
\label{figure5}
\end{figure}

Figure~\ref{figure5} shows a typical diagnostic trace near the beginning of a run 
in the unstable region, revealing significant fluctuations in the densities of 
the coexisting phases.  To analyze these variations, it is helpful to define the 
root-mean-square (rms) fluctuations
\begin{equation}
\sigma_i \equiv \sqrt{\frac{1}{N_{MC}}
\sum_{j=1}^{N_{MC}}\left(\phi_i^{(j)}-\phi_i\right)^2}~,
\label{sigma}
\end{equation}
where $\phi_i$ is the mean volume fraction of species $i$ ($i=c,p$) and
$\phi_i^{(j)}$ is the volume fraction of species $i$ sampled at Monte Carlo step $j$.
\begin{figure}
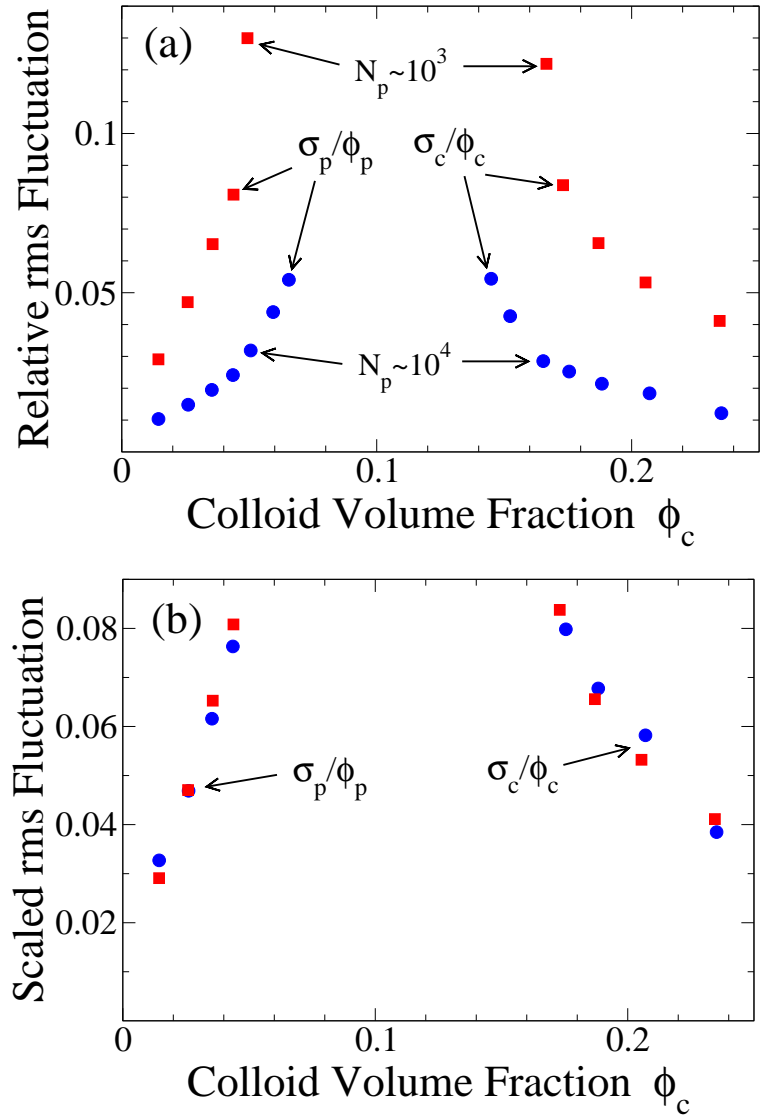

\centering{
\vspace*{1cm}
\includegraphics[width=0.6\textwidth,angle=0]{sigma.eps}
\\[2ex]
\includegraphics[width=0.6\textwidth,angle=0]{sigma-scaled.eps}
}
\caption{Root-mean-square (rms) fluctuations in densities of coexisting phases,
corresponding to the phase diagram in Fig.~\ref{figure4}.
(a) Relative fluctuations $\sigma_c/\phi_c$ and $\sigma_p/\phi_p$ for systems 
with polymer numbers in the range $N_p=6\times 10^2$-$10^3$ (squares) and 
for systems with 10 times as many particles, $N_p=6\times 10^3$-$10^4$ (circles).  
(b) Same data, but with larger-system fluctuations now scaled by a factor of $\sqrt{10}$.
}
\label{figure6}
\end{figure}
Figure~\ref{figure6} shows the relative rms fluctuations, $\sigma_c/\phi_c$
and $\sigma_p/\phi_p$, for two sets of systems -- one relatively small,
with colloid number $N_c=100$ and total polymer numbers in the range 
$N_p=6\times 10^2$-$10^3$, and the other 10 times larger, with $N_c=1000$ and 
$N_p=6\times 10^3$-$10^4$ -- computed from runs of $2\times 10^5$ MC steps.
To minimize correlations between successive samples of the volume fraction, 
we spaced the samples by intervals of 1000 steps.

As seen in panel (a) of Fig.~\ref{figure6}, fluctuations grow upon approaching 
the critical point and as the system size is decreased.
Panel (b) replots the data from panel (a), but with fluctuations in the larger 
systems now scaled by a factor of $\sqrt{10}$ (square-root of system size ratio).
The collapse of the two data sets upon scaling demonstrates the well-known result
that relative fluctuations scale with the inverse-square-root of the particle number:
$\sigma_i/\phi_i\sim 1/\sqrt{N}$.  As an exercise, students could test this 
scaling property by performing simulations over a range of system size.

\section{Concluding Remarks}\label{Conclusions}
In summary, we have extended to fluid mixtures a version of the Gibbs ensemble 
Monte Carlo simulation method that models both phases virtually via a thermodynamic 
equation of state or free energy.  As illustrations, we have applied the method 
to demixing in two models for which, in a mean-field approximation, analytic 
free energy expressions are known: a lattice model of simple mixtures and a 
free-volume model of colloid-polymer mixtures.  For both models, we have validated 
the method by computing fluid demixing phase diagrams that closely agree with 
those calculated from Maxwell common-tangent constructions.

As a computational method, virtual GEMC has two main virtues.  First, for 
finite systems whose equation of state may be known approximately from 
experiment, but whose interparticle interactions remain unknown, the method 
provides an alternative to a Maxwell construction that incorporates the impact 
of fluctuations on demixing.  Second, and perhaps more importantly, 
the virtual GEMC method has pedagogical value as a tool for demonstrating the 
statistical nature of phase behavior and for allowing rapid exploration and 
analysis of fluctuations in finite systems.
Our Java simulation programs can be readily adapted for use as classroom 
demonstrations or as numerical ``experiments" in a computational laboratory.
Finally, it should be straightforward to generalize the virtual GEMC method to 
other models and to multi-component mixtures.  Future applications, for example, 
may explore demixing phase diagrams of ternary oil-water-surfactant mixtures.

\ack
This work was supported by the National Science Foundation under Grant No.~DMR-1106331
and by the American Chemical Society Petroleum Research Fund (PRF 44365-AC7).
Helpful discussions with Ben Lu are gratefully acknowledged.








\vspace*{1cm}
\begin{thebibliography}{99}


\bibitem{chandler}
D.~Chandler, {\it Introduction to Modern Statistical Mechanics} 
(Oxford, Oxford, 1987).

\bibitem{frenkel01}
D.~Frenkel and B.~Smit, {\it Understanding Molecular Simulation},
2nd ed.~(Academic, London, 2001).

\bibitem{allen87}
M.~P.~Allen and D.~J.~Tildesley, {\it Computer Simulation of Liquids}
(Oxford, Oxford, 1987).

\bibitem{landau73}
D.~P.~Landau and R.~Alben, ``Monte Carlo Calculations as an Aid in 
Teaching Statistical Mechanics,"
\AJP {\bf 41}, 394-400 (1973).

\bibitem{sauer81}
G.~Sauer, ``Teaching classical statistical mechanics: A simulation approach,"
\AJP {\bf 49}, 13-19 (1981).

\bibitem{wilding01}
N.~B.~Wilding, ``Computer simulation of fluid phase transitions,"
\AJP {\bf 69}, 1147-1155 (2001).

\bibitem{panagiotopoulos87}
A.~Z.~Panagiotopoulos, ``Direct determination of phase coexistence
properties of fluids by Monte Carlo simulation in a new ensemble,"
\MP {\bf 61}, 813-826 (1987). 

\bibitem{panagiotopoulos88}
A.~Z.~Panagiotopoulos, N.~Quirke, M.~Stapleton, and D.~J.~Tildesley, 
``Phase equilibria by simulation in the Gibbs ensemble: alternative derivation, 
generalization and application to mixture and membrane equilibria,"
\MP {\bf 63}, 527-545 (1988).

\bibitem{panagiotopoulos92}
A.~Z.~Panagiotopoulos, ``Direct Determination of Fluid Phase Equilibria 
by Simulation in the Gibbs Ensemble: A Review,"
{\it Mol. Sim.} {\bf 9}, 1-23 (1992).

\bibitem{panagiotopoulos95}
A.~Z.~Panagiotopoulos, ``Gibbs Ensemble Techniques," 
in {\it Observation, Prediction, and Simulation of Phase Transitions in Complex Fluids},
NATO ASI Series C {\bf 460}, edited by M.~Baus, L.~R.~Rull, and J.~P.~Ryckaert, 
463-501 (Kluwer, Dordrecht, 1995).

\bibitem{bolhuis02}
P.~G.~Bolhuis, A.~A.~Louis, and J.-P.~Hansen,
``Influence of Polymer-Excluded Volume on the Phase-Behavior of 
Colloid-Polymer Mixtures," \PRL {\bf 89}, 128302-1-4 (2002). 

\bibitem{lu-denton11}
B.~Lu and A.~R.~Denton, ``Crowding of polymer coils and demixing in
nanoparticle-polymer mixtures," \JPCM {\bf 23}, 285102-1-9 (2011).

\bibitem{lu-denton07}
B.~Lu and A.~R.~Denton, ``Phase separation of charge-stabilized colloids:
A Gibbs ensemble Monte Carlo simulation study," \PR E {\bf 75} 061403-1-9 (2007).


\bibitem{kofke93}
M.~Mehta and D.~A.~Kofke, ``Implementation of the Gibbs ensemble using a thermodynamic model
for one of the coexisting phases," \MP {\bf 79}, 39-52 (1993).

\bibitem{metropolis53}
N.~Metropolis, A.~W.~Rosenbluth, M.~N.~Rosenbluth, A.~H.~Teller, and E.~Teller,
``Equation of State Calculations by Fast Computing Machines,"
\JCP {\bf 21}, 1087-1092 (1953).

\bibitem{flory69}
P.~J.~Flory, {\it Statistical Mechanics of Chain Molecules}
(Wiley, New York, 1969).

\bibitem{degennes79}
P.~G.~de Gennes, {\it Scaling Concepts in Polymer Physics}
(Cornell, Ithaca, 1979).

\bibitem{osp}
H.~Gould, J.~Tobochnik, and W.~Christian, {\it An Introduction to Computer Simulation 
Methods}, 3rd ed.~(Addison Wesley, 2006).

\bibitem{download}
Our Java simulation programs, implementing the virtual GEMC method within the 
Open Source Physics library, are freely available upon request.

\bibitem{pusey}
P.~N.~Pusey, in {\it Liquids, Freezing and Glass Transition},
session 51, ed. J.-P.~Hansen, D.~Levesque, and J.~Zinn-Justin
(North-Holland, Amsterdam, 1991).

\bibitem{ilett95}
S.~M.~Ilett, A.~Orrock, W.~C.~K.~Poon, and P.~N.~Pusey, ``Phase behavior of
a model colloid-polymer mixture," \PR E {\bf 51}, 1344-1352 (1995).

\bibitem{poon02}
W.~C.~K.~Poon, ``The physics of a model colloid-polymer mixture," 
\JPCM {\bf 14}, R859-R880 (2002).

\bibitem{ao54}
S.~Asakura and F.~Oosawa, ``Surface Tension of High‐Polymer Solutions,"
\JCP {\bf 22}, 1255 (1954). 

\bibitem{vrij76}
A.~Vrij, ``Polymers at interfaces and the interactions in colloidal
dispersions, {\it Pure \& Appl. Chem.} {\bf 48}, 471-483 (1976).

\bibitem{denton-schmidt02}
A.~R.~Denton and M.~Schmidt, \JPCM {\bf 14}, 12051 (2002).

\bibitem{lekkerkerker92}
H.~N.~W.~Lekkerkerker, W.~C.~K.~Poon, P.~N.~Pusey, A.~Stroobants, and P.~B.~Warren,
\EPL {\bf 20}, 559 (1992).

\bibitem{roth06}
S.~M.~Oversteegen and R.~Roth, \JCP {\bf 122}, 214502 (2005). 

\bibitem{hansen90}
J.-P.~Hansen and I.~R.~McDonald, {\it Theory of Simple Liquids},
$2 ^{nd}$ ed.~(Academic, London, 1986).

\bibitem{kofke94}
M.~Mehta and D.~A.~Kofke, ``Coexistence diagrams of mixtures by molecular simulation,"
{\it Chem. Eng. Sci.} {\bf 49}, 2633-2645 (1994).


\end{thebibliography}
\end{document}